\documentclass[aps,prb,reprint,superscriptaddress]{revtex4-2}

\usepackage{amsmath}
\usepackage{mathrsfs}
\usepackage{amssymb}
\usepackage{graphicx}
\usepackage{bm}% bold maths
\usepackage{enumerate}
\usepackage[colorlinks=true,citecolor=blue,linkcolor=blue]{hyperref}
\usepackage{siunitx}

\begin{document}

%Title of paper
\title{Large Bilinear Magnetoresistance from Rashba Spin-Splitting on the Surface of a Topological Insulator}

\author{Yang Wang}
\thanks{These authors contributed equally}
%\email[]{yangwang@udel.edu}
%\altaffiliation{}
\affiliation{Department of Physics and Astronomy, University of Delaware, Newark, Delaware, 19716, USA}

\author{Binbin Liu}
\thanks{These authors contributed equally}
\affiliation{School of Physics, Beihang University, Beijing 100191, China}

\author{Yue-Xin Huang}
\thanks{These authors contributed equally}
\affiliation{Research Laboratory for Quantum Materials, Singapore University of Technology and Design, Singapore 487372, Singapore}

\author{Sivakumar V. Mambakkam}
\affiliation{Department of Materials Science and Engineering, University of Delaware, Newark, Delaware, 19716, USA}

\author{Yong Wang}
\affiliation{Department of Materials Science and Engineering, University of Delaware, Newark, Delaware, 19716, USA}

\author{Shengyuan A. Yang}
\affiliation{Research Laboratory for Quantum Materials, Singapore University of Technology and Design, Singapore 487372, Singapore}

\author{Xian-Lei Sheng}
\affiliation{School of Physics, Beihang University, Beijing 100191, China}

\author{Stephanie A. Law}
\affiliation{Department of Physics and Astronomy, University of Delaware, Newark, Delaware, 19716, USA}
\affiliation{Department of Materials Science and Engineering, University of Delaware, Newark, Delaware, 19716, USA}

\author{John Q. Xiao}
\email[]{jqx@udel.edu}
\affiliation{Department of Physics and Astronomy, University of Delaware, Newark, Delaware, 19716, USA}

\date{\today}

\begin{abstract}
In addition to the topologically protected linear dispersion, a band-bending-confined two-dimensional electron gas with tunable Rashba spin-splitting (RSS) was found to coexist with the topological surface states on the surface of topological insulators (TIs). Here, we report the observation of large bilinear magnetoresistance (BMR) in Bi\textsubscript{2}Se\textsubscript{3} films decorated with transition metal atoms. The magnitude of the BMR sensitively depends on the type and amount of atoms deposited, with a maximum achieved value close to those of strong Rashba semiconductors. Our first-principles calculations reproduce the quantum well states and reveal sizable RSS in all Bi\textsubscript{2}Se\textsubscript{3} heterostructures with broken inversion symmetry. Our results show that charge-spin interconversion through RSS states in TIs can be fine-tuned through surface atom deposition and easily detected via BMR for potential spintronic applications.
\end{abstract}

% insert suggested keywords - APS authors don't need to do this
%\keywords{}

%\maketitle must follow title, authors, abstract, and keywords
\maketitle

Bilinear or unidirectional magnetoresistance in nonmagnetic systems without inversion symmetry \cite{Tokura,Ideue,He1,Guillet,He2,Y.Wang,Legg} describes the difference in resistance when the electric current or magnetic field direction is switched. It can be phenomenologically expressed as \cite{Tokura,Ideue,Rikken1}
\begin{eqnarray}
	R(I,B)=R_0[1+\beta B^2+\gamma \bm I\cdot (\bm P \times \bm B)]
	\label{Eq1}
\end{eqnarray}
where $R_0$ is the resistance at zero magnetic field, $\bm I$ is the electric current, $\bm B$ is the magnetic field, and $\bm P$ represents the polar direction of the conductor. The second term describes the normal magnetoresistance. The magnitude of the BMR is measured by the coefficient $\gamma$. Such rectification effect typically exists in chiral conductors \cite{Rikken2,Pop} or systems where the spin-degeneracy is lifted by inversion symmetry breaking in combination with spin-orbit coupling (SOC) \cite{Ideue,He1,Guillet,He2}. For the latter case, topological insulator surfaces \cite{Kane,S.C.Zhang} and Rashba spin-splitting \cite{Rashba,Bychkov} states with spin-momentum locked spin-textures are canonical examples.

\begin{figure*}[]
	\includegraphics[width=0.95\linewidth]{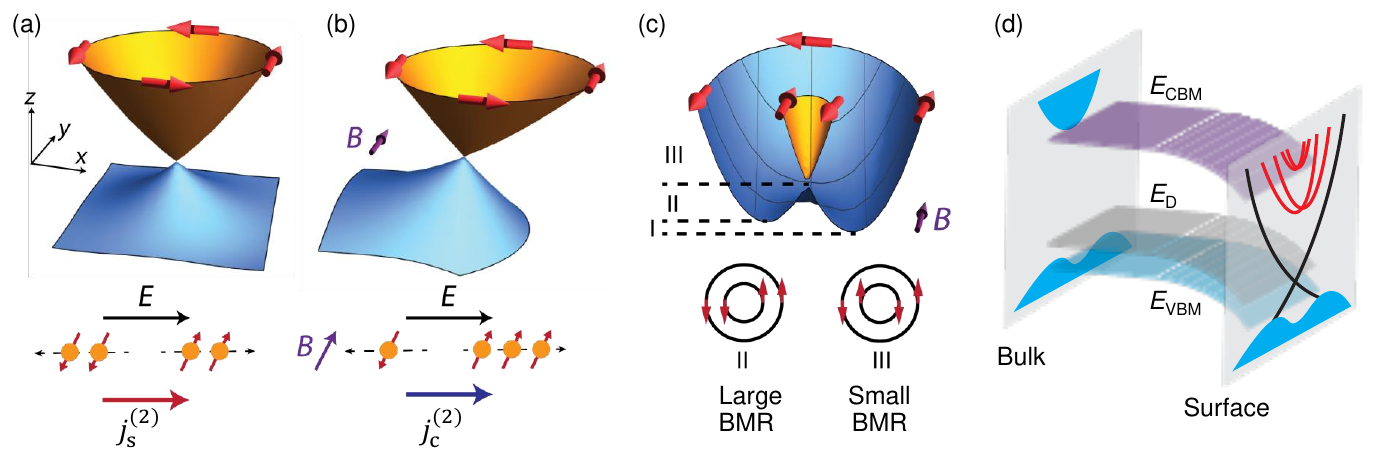}
	\caption{Illustration of the BMR and band-bending mechanisms. (a) Driven by an $x$-direction electric field, there is a second-order spin current generated in the states of an inner Rashba band or the surface TI dispersion with particle-hole asymmetry, due to spin-momentum locking. (b) When a magnetic field is applied in the $y$-direction, it distorts the Fermi contour asymmetrically and causes an imbalance between the right- and left-moving electrons with opposite spin polarizations, resulting in a net second-order charge current. (c) The cross-section of a complete Rashba band under an in-plane mangetic field. Regions II and III have large and small BMR responses, respectively. Region I is narrow and can be neglected. (d) From the TI bulk to surface, the band bends downward and a ladder of QW states are formed in the surface 2DEG. When there is also substantial electric potential gradient, these QW states are spin-split by Rashba SOC. $E_\text{CBM}$, $E_\text{D}$, and $E_\text{VBM}$ denote the energy of conduction band minimum, Dirac point, and valence band maximum, respectively.}
	\label{Fig1}
\end{figure*}

As illustrated in Figs.~\ref{Fig1}(a) and~\ref{Fig1}(b), taking the TI surface dispersion or the inner half of a Rashba-type band as an example, when driven by an $x$-direction electric field, a second-order spin current is generated through spin-momentum locking \cite{Hamamoto}. It can be intuitively understood as equal amounts of electrons with opposite spin polarizations moving in opposite directions, so there is a nonlinear spin but no charge current. When a magnetic field is applied, the upright Dirac cone is sheared due to the existence of nonlinear-in-momentum terms arising from e.g., particle-hole asymmetry ($k^2$) or hexagonal warping ($k^3$). This causes imbalance between the left- and right-moving electrons, and the nonlinear spin current is partially converted into a nonlinear charge current, giving rise to the BMR \cite{He1,S.L.Zhang,He3,Li}. As to the full Rashba band [Fig.~\ref{Fig1}(c)], when the Fermi level is above the charge neutral point (CNP) (region III), due to the cancellation of the two Fermi contours with opposite spin helicities, the BMR is small. When the Fermi level lies in region II, a large BMR arises from the addition of the two Fermi contours \cite{Ideue,Li}. There is also a narrow region I induced by Zeeman splitting which also exhibits large BMR. However, for the small magnetic fields (~0.2 T) used in this study, this region of width $\sim$0.1 meV can be neglected.

Experimentally, a BMR with a magnitude of $\gamma\sim$0.01 \si{A^{-1}B^{-1}} was reported in Bi\textsubscript{2}Se\textsubscript{3} grown on Al\textsubscript{2}O\textsubscript{3} substrates \cite{He1}. It was attributed to the hexagonal warping effect in the topological surface states (TSS). Much larger BMR was achieved through the RSS mechanism [Fig.~\ref{Fig1}(c)], with $\gamma$ reaching $\sim$1 \si{A^{-1}T^{-1}} in the polar semiconductor BiTeBr \cite{Ideue} and $\sim$10-100 \si{A^{-1}T^{-1}} in the two-dimensional electron gases (2DEGs) formed at oxide interfaces \cite{He2,Choe} Actually, as sketched in Fig.~\ref{Fig1}(d), early angle-resolved photoelectron spectroscopy (ARPES) and computational studies revealed that on the surface of Bi\textsubscript{2}Se\textsubscript{3} a 2DEG is confined in the top most several QLs due to band-bending (BB)\cite{Bianchi,Bahramy}. Further introduction of an electrostatic potential gradient either by electric gating\cite{King} or surface atoms evaporation\cite{Zhu} will cause tunable and robust RSS in these QW states. It is natural to ask, can these spin-split states on the surface of TIs generate large BMR?

In this study, we observed large tunable BMR in thick Bi\textsubscript{2}Se\textsubscript{3} films decorated with transition metal (TM) atoms. The maximum nonreciprocal coefficient $\gamma$ achieved is comparable to these of strong Rashba materials. Our density-functional theory (DFT) calculations reproduce the ladder of QW states with sizable RSS in all inversion asymmetric Bi\textsubscript{2}Se\textsubscript{3} heterostructures. Through analysis of the temperature dependence, we deduce the large BMR is mainly from the states in region II of the Rashba band, accessed during the band-bending process. As a complement to previous ARPES measurements, our work further validates TIs from nonreciprocal transport aspects as a highly tunable Rashba material to explore charge-spin interconversion phenomena through nontopological bands.

\begin{figure*}[]
	\includegraphics[width=1\linewidth]{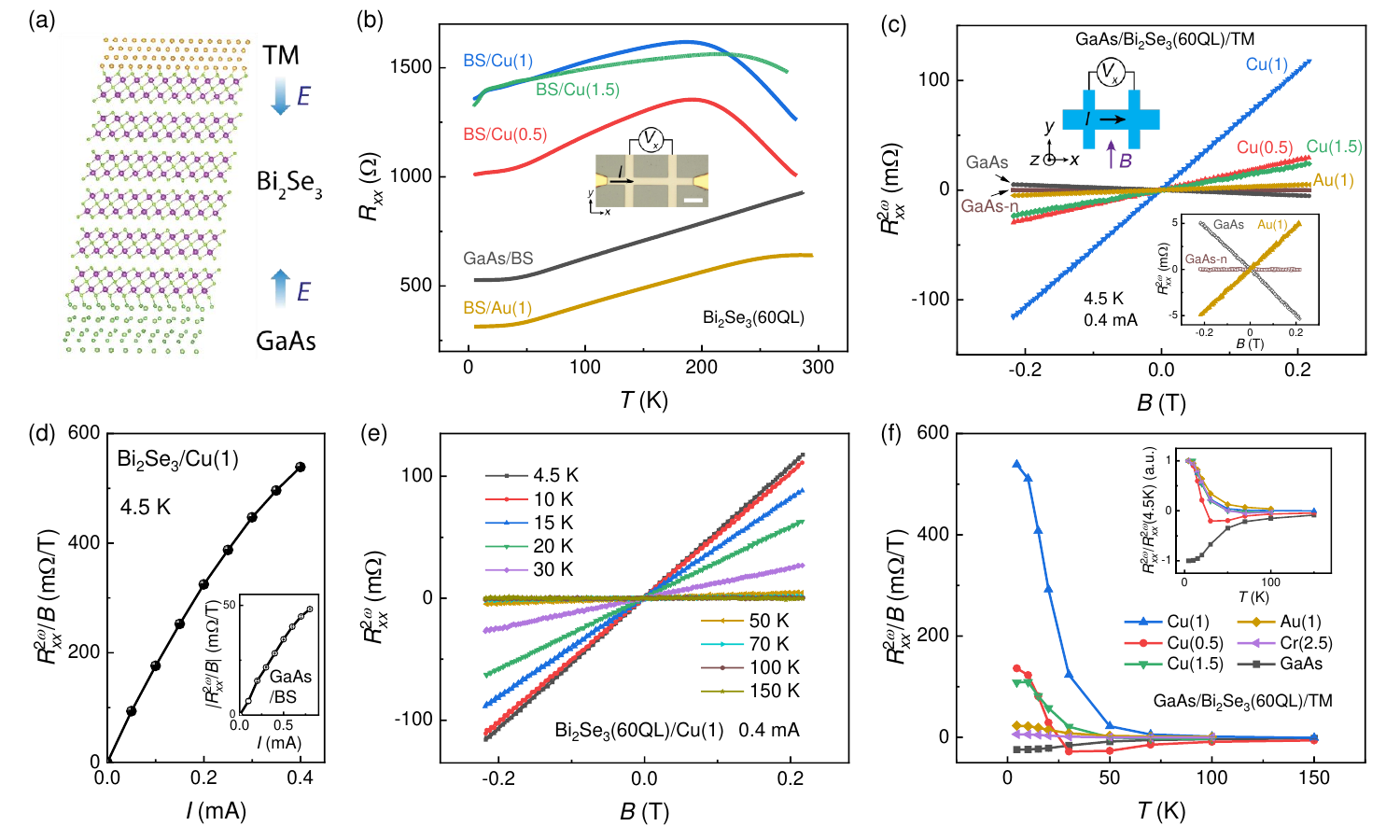}
	\caption{(a) Crystalline structure of the GaAs/Bi\textsubscript{2}Se\textsubscript{3}/TM heterostructure. The blue arrows represent electric field potential induced by the GaAs substrate or deposited TM. (b) Longitudinal resistance versus temperature for various Bi\textsubscript{2}Se\textsubscript{3} heterostructures. The inset depicts a device image with scale bar 20 $\mu$m. (c) Longitudinal second-harmonic resistance as a function of $y$-direction magnetic field in various devices. The top inset illustrates the measurement setup. The bottom inset is a zoomed-in plot of the GaAs/$n$-BS, GaAs/BS, and BS/Au(1) samples. (d) The slope  $R_{xx}^{2\omega}/B$ exhibits a linear dependence on current. (e) $R_{xx}^{2\omega}$ vs $B$ of the BS/Cu(1) sample at different temperatures. (f) Summary of the temperature dependence of the BMR magnitude in different devices. Inset is normalized BMR magnitude.}
	\label{Fig2}
\end{figure*}

The 60 quintuple-layer (QL) Bi\textsubscript{2}Se\textsubscript{3} (BS) films used in this study were grown on GaAs substrates by the molecular beam epitaxy (MBE) method and decorated with 0.5-1.5 nm thick Cu or Au atoms (See Supplemental Material \cite{Suppl} for growth and characterization details). The Hall bar devices were patterned by the standard photolithography method. Because of the small amount of metal deposited, in transport measurements the current essentially only goes through the 60 QL BS layer. As depicted in Fig.~\ref{Fig2}(a), both the substrate and deposited TM atoms can break the inversion symmetry of Bi\textsubscript{2}Se\textsubscript{3}, introduce an electric potential gradient, and cause substantial RSS. Additionally, the Cu and Au atoms migrate into the van der Waals layered Bi\textsubscript{2}Se\textsubscript{3} structure \cite{Y.L.Wang,S.J.Chang} and $p$- or $n$-dope it. The doping effects can be seen in the linear transport regime. As displayed in Fig.~\ref{Fig2}(b), compared with bare BS film grown on GaAs, Cu or Au raises or lowers the resistance of BS, suggesting that they are working as electron acceptors or donors, respectively. Interestingly, although the BS/Au sample maintains metallic behavior, the resistance of BS/Cu devices first increases and then decreases when temperature is lowered from 290 K to 5 K, with a peak around 200 K. Such nonmonotonic behavior was also observed in Cu-doped Bi\textsubscript{2}Te\textsubscript{3} and was attributed to the change in carrier density \cite{H.J.Wu,Dan}. Moreover, the smaller resistance in BS/Cu(0.5) and the similar resistances in BS/Cu(1 and 1.5) samples suggest that saturation of the doping effect occurs between deposition of 0.5 to 1 nm Cu. From Hall measurements, the sheet carrier density $n_\text{s}$ for the GaAs/BS, BS/Cu(0.5, 1, and 1.5), and BS/Au(1) samples are 5.2, 3.9, 4.1, 5.1, and 5.6$\times10^{13}$ \si{cm^{-2}}, respectively. These results suggest that Cu doping can suppress the bulk conduction for the as-grown $n$-doped Bi\textsubscript{2}Se\textsubscript{3} films.

The effect from electrostatic potential on Bi\textsubscript{2}Se\textsubscript{3} is revealed by nonreciprocal transport responses, which sensitively depends on the symmetry and band structure of the material \cite{Tokura}. As indicated by the name, the bilinear magnetoresistance depends linearly on both external magnetic and electric fields. As sketched in Fig.~\ref{Fig2}(c) top inset, we sent a low frequency a.c. current in the $x$-direction, swept the magnetic field in the $y$-direction, and measured the longitudinal second-harmonic resistance (SHR) $R_{xx}^{2\omega}\equiv V_{x}^{2\omega}/I$ by the standard lock-in technique. $R_{xx}^{2\omega}$ in all samples exhibits linear dependence on $B$, as expected. GaAs-$n$ denotes another Bi\textsubscript{2}Se\textsubscript{3} (60 QL) sample also grown on GaAs substrate, but it has a much higher carrier density of $2.7\times10^{14}$ \si{cm^{-2}}, so the current is shunt through the bulk. The SHR of it is negligibly small, showing that the inversion-symmetric BS bulk does not contribute to the BMR. BMR is detected in other samples where substantial current is carried by the surface states. As shown in Fig.~\ref{Fig2}(c) bottom inset, the BMR in GaAs/BS and BS/Au(1) have opposite signs, corresponding to the bottom and top surface contributions, respectively. The BMR magnitude $\gamma$ in the GaAs/BS sample already reaches 0.08 \si{A^{-1}T^{-1}}, which is about an order of magnitude larger than that in Al\textsubscript{2}O\textsubscript{3}/BS samples with similar carrier densities \cite{He1}. This suggests spin-split QW states instead of hexagonally warped TSS as the origin of the observed BMR. The reason why RSS-induced BMR was not dominant in Al\textsubscript{2}O\textsubscript{3}/BS is possibly because the downward band-bending effect is weak in the thin BS(20 QL) film grown on the high-$k$ dielectric Al\textsubscript{2}O\textsubscript{3} substrate \cite{Steinberg}. The BMR can be further enhanced by the deposition of TM atoms. In the BS/Cu(1) sample, $\gamma$ reaches 2.0 \si{A^{-1}T^{-1}}, surpassing that of the strong Rashba semiconductor BiTeBr ($\gamma \sim 1$) \cite{Ideue}. We noticed that compared with Ref. \cite{Ideue,He1}, the SHR in our samples is large and of low noise. This is rare in 3D polar conductors, suggesting the robustness of the electrostatic-potential-induced RSS on TI surfaces. Fig.~\ref{Fig2}(d) shows that the slope $R_{xx}^{2\omega}/B$ also scales linearly with current, together with Fig.~\ref{Fig2}(c) demonstrating the bilinear nature of the measured resistance. The slight deviation from linear dependence under large currents is due to Joule heating.

We confirmed that that under an $x$-direction field, the longitudinal SHR becomes much smaller, consistent with the selection rule in Eq.~\eqref{Eq1}. We also observed the transverse counterpart of BMR, the nonlinear planar Hall effect \cite{He3} in our samples. (See Supplemental Note 4 for details.) As represented by the BS/Cu(1) sample [Fig.~\ref{Fig2}(e)], the magnitude of the BMR decreases monotonically from being maximum at 4.5 K to being negligible at 70 K. This trend is observed in all the samples [Fig.~\ref{Fig2}(f)]. Moreover, as shown in the inset of Fig.~\ref{Fig2}(f), except the BS/Cu(0.5) sample, the normalized BMR magnitude $R_{xx}^{2\omega}(T)/R_{xx}^{2\omega}(4.5~\si{K})$ exhibits very similar temperature dependence in different samples. For the BS/Cu(0.5) sample, we suspect that the electrostatic perturbation is relatively weak due to the tiny amount (0.5 nm) of Cu deposited. This may cause the BMR from Bi\textsubscript{2}Se\textsubscript{3} top surface to have a faster decay trend as temperature increases compared to that of the bottom surface. As a result of the competition between the top and bottom surfaces, a sign reversal of $R_{xx}^{2\omega}/B$ around 25 K is observed.

\begin{figure*}[]
	\includegraphics[width=1\linewidth]{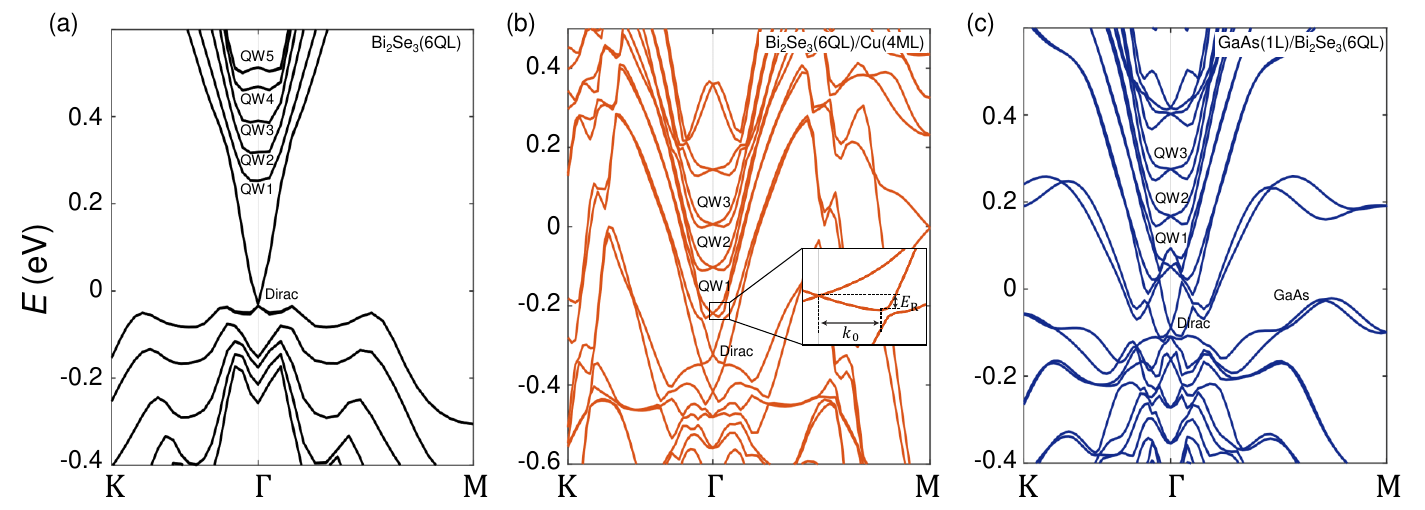}
	\caption{Calculated band structures of (a) Bi\textsubscript{2}Se\textsubscript{3}(6QL), (b) Bi\textsubscript{2}Se\textsubscript{3}(6QL)/Cu(4ML), and (c) GaAs(1L)/Bi\textsubscript{2}Se\textsubscript{3}(6QL). Dirac and QW denote the topological surface and nontopological quantum well states, respectively. The inset in (b) is a zoom-in image of the QW1 states, with the momentum offset $k_0$ and energy splitting $E_R$ marked in the figure.}
	\label{Fig3}
\end{figure*}

To support our claim that the observed BMR is from RSS, we carried out density-functional theory (DFT) calculations for Bi\textsubscript{2}Se\textsubscript{3} heterostructures made with 6 QL Bi\textsubscript{2}Se\textsubscript{3} and 1-4 monolayer (ML) Cu, 2 ML Au or 1 layer (L) GaAs. Because the 2DEG on Bi\textsubscript{2}Se\textsubscript{3} is confined in the top few ($\sim$10) QLs due to band bending \cite{Bahramy}, which gives its unique quantum well (QW) states other than the continuous bulk bands, we believe the calculations on 6 QL Bi\textsubscript{2}Se\textsubscript{3} can capture the essential features of the real 2DEG states formed on the surface of our Bi\textsubscript{2}Se\textsubscript{3} samples. The calculated band structures are shown in Fig.~\ref{Fig3}. In the standalone Bi\textsubscript{2}Se\textsubscript{3} film with preserved inversion symmetry, besides the topologically protected gapless surface Dirac dispersion, the bulk states are quantized into a ladder of QW states [Fig.~\ref{Fig3}(a)]. All the states are spin-degenerate in momentum space, and BMR is correspondingly forbidden. As shown in Fig.~\ref{Fig3}(b) and~\ref{Fig3}(c), when Bi\textsubscript{2}Se\textsubscript{3} is proximately coupled to either Cu or GaAs, inversion symmetry breaking together with SOC lifts the spin degeneracy and causes Rashba-type spin-splitting in all the QW states. The size of RSS is described by the Rashba coefficient $\alpha_R=2E_\text{R}/k_0$ , where $E_\text{R}$ and $k_0$ are the energy splitting and momentum offset, respectively, as defined in the inset of Fig.~\ref{Fig3}(b). The BS/Cu slab has well-separated topological surface and nontopological QW states with sizable RSS. While in GaAs/BS, the GaAs bands hybridize with BS bands and unexpectedly large RSS (with maximum 2.3 \si{eV\AA}) is observed at low QW states possibly due to the band anti-crossing features \cite{Acosta}. However, in real samples, the bottom and top BS surfaces have different interfacial qualities, so it is reasonable to only compare the RSS at the top BS/Cu or Au interfaces.

\begin{figure*}[]
	\includegraphics[width=1\linewidth]{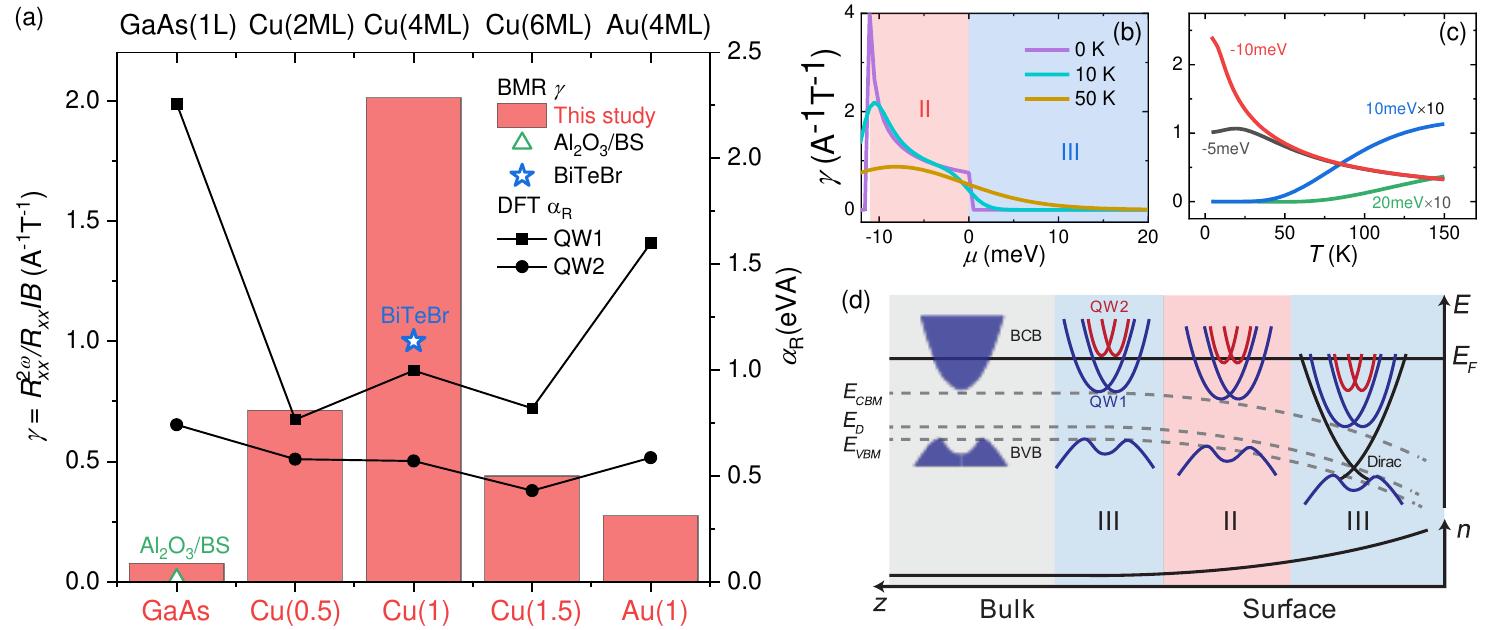}
	\caption{(a) Experimentally measured BMR magnitude $\gamma$ and calculated Rashba coefficient $\alpha_R$ for various Bi\textsubscript{2}Se\textsubscript{3} heterostructures. The triangle and star denote the  $\gamma$ values for Al\textsubscript{2}O\textsubscript{3}/Bi\textsubscript{2}Se\textsubscript{3} (\cite{He1}) and BiTeBr (\cite{Ideue}), respectively. (b) and (c) Calculated chemical potential (b) and temperature (c) dependence of the BMR magnitude. (d) Illustration of the band-bending process. BCB and BVB denote bulk conduction and valence bands, respectively. CBM/VBM is the conduction/valence band minimum/maximum. Going from bulk to surface, the quantum well states gradually form, and the Fermi level crosses region III and II during the BB process. It is the states in region II that contribute to the large BMR.}
	\label{Fig4}
\end{figure*}

As summarized in Fig.~\ref{Fig4}(a), QW1 has the largest $\alpha_\text{R}$ in each material, reaching 1 and 1.6 \si{eV\AA} in the BS/Cu(4ML) and BS/Au(4ML) slabs, respectively. Importantly, the higher QW states (QW2, 3, …) also exhibit sizable RSS, at the order of ~0.5 \si{eV\AA}. These features are consistent with previous ARPES \cite{Zhu} and DFT \cite{H.Yang} results, and are crucial for the observation of the BMR on the surface of TIs, as discussed below. Fig.~\ref{Fig4}(a) also displays the BMR magnitude $\gamma$ of the Bi\textsubscript{2}Se\textsubscript{3} heterostructures made in this study together with those of Al\textsubscript{2}O\textsubscript{3}/Bi\textsubscript{2}Se\textsubscript{3} \cite{He1} and BiTeBr \cite{Ideue}. Besides exhibiting large values, we find that $\gamma$ does not follow the same trend as that of $\alpha_\text{R}$ in QW1 or QW2. This means that RSS is a necessary but not sufficient condition for the observation of large BMR. There are other factors that control the magnitude of BMR. Different from the perfect lattice structure used in the calculation, the MBE-grown GaAs/BS has an imperfect interface with reduced SOC strength, and correspondingly, the real RSS should be smaller than predicted. Besides, defects formation also makes the bottom surface more resistive, so less amount of current goes through it as compared to the top surface. This can explain the small BMR observed in the GaAs/BS sample. Similarly, in the BS/Au sample, majority of the current is shunted through the heavily $n$-doped bulk, so despite having a larger RSS, its BMR magnitude is smaller than that of BS/Cu samples.

These above considerations cannot explain the significant drop of $\gamma$ from 2 \si{A^{-1}T^{-1}} in BS/Cu(1) to 0.4 \si{A^{-1}T^{-1}} in BS/Cu(1.5), given that these two samples have very similar resistivities [Fig.~\ref{Fig2}(b)] and $\alpha_\text{R}$ [Fig.~\ref{Fig4}(a)]. Here we propose a band-bending picture to explain this discrepancy. The model Hamiltonian of a Rashba band under a $y$-direction magnetic field can be written as
\begin{eqnarray}
	H= \hbar^2 k^2/{2m^*}+\alpha_\text{R}(k_x \sigma_y - k_y\sigma_x)
	+B_y \sigma_y,
	\label{Eq2}
\end{eqnarray}
where $\hbar$ is the reduced Planck constant, $m^*$ is the effective mass of the electrons, $\bm\sigma$ represent Pauli spin matrices, and $B_y$ is the magnetic field measured in energy unit. Using semiclassical Boltzmann approach, we numerically calculated the chemical potential and temperature dependence of the BMR (See Supplemental Note 5 for details). As plotted in Fig.~\ref{Fig4}(b), consistent with the qualitative argument in the introduction [Fig.~\ref{Fig1}(c)], at $T=0$ the BMR is zero when chemical potential $\mu$ lies in region III and only becomes finite when temperature increases or $\mu$ goes into region II. Fig.~\ref{Fig4}(c) shows that in region II $\gamma$ decreases when temperature is increased, which is consistent with our experimental results [Fig.~\ref{Fig2}(f)]. This together with the opposite temperature dependence of the BMR and its small magnitude in region III suggest that the observed BMR is dominantly from region II. Based on the measured carrier density $n_\text{s}$ ($3.9-5.6\times10^{13}$ \si{cm^{-2}}), the bulk Fermi level relative to the unbent surface Dirac point is $\sim$0.3 eV, which is much higher than the level of the bulk conduction band minimum (CBM) for Bi\textsubscript{2}Se\textsubscript{3} ($\sim$0.2 eV) \cite{Bianchi,Zhu}. Given the energy splitting $E_\text{R}\sim10$ \si{meV}, region II of QW1 cannot be accessed during the downward BB process. This leads us to draw the schematic picture as shown in Fig.~\ref{Fig4}(d). Different from the topological surface states, the QW states of the 2DEG on the TI surface spread over a depth of $\sim$10 QL \cite{Bahramy}. Going from bulk to surface, during the BB process, the position of the Fermi level relative to the QW1 minimum shifts up gradually, and the region II of higher QW levels (e.g., QW2) is accessed. It is these states that make dominant contributions to the observed large BMR which weakens as temperature increases. As sketched at the bottom of Fig.~\ref{Fig4}(d), due to the exponential carrier density profile of the 2DEG \cite{Bianchi,King}, the number of electrons in region II takes a maximum if the local Fermi level at the surface happens to be in region II. This requires a delicate balance between the bulk Fermi level and the BB strength. This can explain the significantly different $\gamma$ in the BS/Cu(1) and BS/Cu(1.5) devices.

In conclusion, we observed large BMR in Bi\textsubscript{2}Se\textsubscript{3} heterostructures surpassing that of strong bulk Rashba semiconductors. On the one hand, we find Cu is a preferential material to engineer Bi\textsubscript{2}Se\textsubscript{3} into a Rashba material, because it not only imposes electrostatic potential to generate sizable RSS, but also suppresses the bulk conduction through diffusive doping. On the other hand, analysis on the temperature dependence suggests that the observed BMR is dominantly from the states below the band-crossing point of higher-level QW states, accessed during the gradual band-bending process. This ensures the easy electrical detection of the RSS through BMR and suggests that BMR should have robust tunability through chemical potential or band-bending engineering. Our work demonstrates BMR as an electrical transport signature to assist the search for inversion asymmetric TI heterostructures with large RSS.

\begin{acknowledgments}
	We thank Yi Ji for the use of the home-built cryogenic transport measurement system. This work was supported by the U.S. DOE, Office of Basic Energy Sciences under Contract No. DE-SC0016380 and by NSF DMR Grant No. 1904076. B. Liu and X.-L. Sheng acknowledge the support from National Natural Science Foundation of China under Grant No. 12174018. Y.-X. H and S. A. Yang are supported by Singapore NRF CRP22-2019-0061. The authors acknowledge the use of the Materials Growth Facility (MGF) at the University of Delaware, which is partially supported by the National Science Foundation Major Research Instrumentation under Grant No. 1828141 and UD-CHARM, a National Science Foundation MRSEC, under Award No. DMR-2011824.
\end{acknowledgments}

% Create the reference section using BibTeX:
%\bibliography{basename of .bib file}

\begin{thebibliography} {}
	\bibitem{Tokura} Y. Tokura and N. Nagaosa, Nat. Commun. {\bf 9}, 3740 (2018).	
	\bibitem{Ideue} T. Ideue, K. Hamamoto, S. Koshikawa, M. Ezawa, S. Shimizu, Y. Kaneko, Y. Tokura, N. Nagaosa, and Y. Iwasa, Nat. Phys. {\bf 13}, 578 (2017).
	\bibitem{He1} P. He, S. S. L. Zhang, D. Zhu, Y. Liu, Y. Wang, J. Yu, G. Vignale, and H. Yang, Nat. Phys. {\bf 14}, 495 (2018).
	\bibitem{Guillet}	T. Guillet, C. Zucchetti, Q. Barbedienne, A. Marty, G. Isella, L. Cagnon, C. Vergnaud, H. Jaffr\`es, N. Reyren, J.-M. George, A. Fert, and M. Jamet, Phys. Rev. Lett. {\bf 124}, 027201 (2020).
	\bibitem{He2}	P. He, S. M. Walker, S. S.-L. Zhang, F. Y. Bruno, M. S. Bahramy, J. M. Lee, R. Ramaswamy, K. Cai, O. Heinonen, G. Vignale, F. Baumberger, and H. Yang, Phys. Rev. Lett. {\bf 120}, 266802 (2018).
	\bibitem{Y.Wang}	Y. Wang, H. F. Legg, T. B\"omerich, J. Park, S. Biesenkamp, A. A. Taskin, M. Braden, A. Rosch, and Y. Ando, Phys. Rev. Lett. {\bf 128}, 176602 (2022).
	\bibitem{Legg} H. F. Legg, M. R\"o{\ss}ler, F. M\"unning, D. Fan, O. Breunig, A. Bliesener, G. Lippertz, A. Uday, A. A. Taskin, D. Loss, J. Klinovaja, and Y. Ando, Nat. Nanotechnol. {\bf 17}, 696 (2022).	
	\bibitem{Rikken1}	G. L. J. A. Rikken and P. Wyder, Phys. Rev. Lett. {\bf 94}, 016601 (2005).
	\bibitem{Rikken2}	G. L. J. A. Rikken, J. F\"olling, and P. Wyder, Phys. Rev. Lett. {\bf 87}, 236602 (2001).
	\bibitem{Pop} F. Pop, P. Auban-Senzier, E. Canadell, G. L. J. A. Rikken, and N. Avarvari, Nat. Commun. {\bf 5}, 3757 (2014).
	\bibitem{Kane}	M. Z. Hasan and C. L. Kane, Rev. Mod. Phys. {\bf 82}, 3045 (2010).
	\bibitem{S.C.Zhang}	X.-L. Qi and S.-C. Zhang, Rev. Mod. Phys. {\bf 83}, 1057 (2011).
	\bibitem{Rashba}	E. I. Rashba, Sov. Phys. Solid State {\bf 2}, 1109 (1960).
	\bibitem{Bychkov}	Y. A. Bychkov and E. I. Rashba, JETP Lett. {\bf 39}, 78 (1984).
	\bibitem{Hamamoto} K. Hamamoto, M. Ezawa, K. W. Kim, T. Morimoto, and N. Nagaosa, Phys. Rev. B {\bf 95}, 224430  (2017).
	\bibitem{S.L.Zhang} S. S.-L. Zhang and G. Vignale, Proc. SPIE {\bf 10732}, 1073215 (2018).
	\bibitem{He3} P. He, S. S.-L. Zhang, D. Zhu, S. Shi, O. G. Heinonen, G. Vignale, and H. Yang, Phys. Rev. Lett. {\bf 123}, 016801 (2019).
	\bibitem{Li} Y. Li, Y. Li, P. Li, B. Fang, X. Yang, Y. Wen, D.-x. Zheng, C.-h. Zhang, X. He, A. Manchon, Z.-H. Cheng, and X.-x. Zhang, Nat. Commun. {\bf 12}, 540 (2021).
	\bibitem{Choe} D. Choe, M.-J. Jin, S.-I. Kim, H.-J. Choi, J. Jo, I. Oh, J. Park, H. Jin, H. C. Koo, B.-C. Min, S. Hong, H.-W. Lee, S.-H. Baek, and J.-W. Yoo, Nat. Commun. {\bf 10}, 4510 (2019).
	\bibitem{Bianchi} M. Bianchi, D. Guan, S. Bao, J. Mi, B. B. Iversen, P. D. C. King, and P. Hofmann, Nat. Commun. {\bf 1}, 128 (2010).
	\bibitem{Bahramy} M. S. Bahramy, P. D. C. King, A. de la Torre, J. Chang, M. Shi, L. Patthey, G. Balakrishnan, Ph. Hofmann, R. Arita, N. Nagaosa, and F. Baumberger, Nat. Commun. {\bf 3}, 1159 (2012).	
	\bibitem{King} P. D. C. King, R. C. Hatch, M. Bianchi, R. Ovsyannikov, C. Lupulescu, G. Landolt, B. Slomski, J. H. Dil, D. Guan, J. L. Mi, E. D. L. Rienks, J. Fink, A. Lindblad, S. Svensson, S. Bao, G. Balakrishnan, B. B. Iversen, J. Osterwalder, W. Eberhardt, F. Baumberger, and P. Hofmann, Phys. Rev. Lett. {\bf 107}, 096802 (2011).
	\bibitem{Zhu} Z.-H. Zhu, G. Levy, B. Ludbrook, C. N. Veenstra, J. A. Rosen, R. Comin, D. Wong, P. Dosanjh, A. Ubaldini, P. Syers, N. P. Butch, J. Paglione, I. S. Elfimov, and A. Damascelli, Phys. Rev. Lett. {\bf 107}, 186405 (2011).
	\bibitem{Suppl} See Supplemental Material at [URL will be inserted by publisher] for details on experimental and computational methods, Bi\textsubscript{2}Se\textsubscript{3} film characterization, the transverse nonlinear Hall measurement results, and model calculations, which includes Refs. \cite{Z.Wang,Kresse1,Kresse2,Blochl,Perdew,Grimme1,Grimme2,Wolos,T.H.Wang}
	\bibitem{Y.L.Wang} Y.-L. Wang, Y. Xu, Y.-P. Jiang, J.-W. Liu, C.-Z. Chang, M. Chen, Z. Li, C.-L. Song, L.-L. Wang, K. He, X. Chen, W.-H. Duan, Q.-K. Xue, and X. C. Ma, Phys. Rev. B {\bf 84}, 075335 (2011).
	\bibitem{S.J.Chang} S.-J. Chang, P.-Y. Chuang, C.-W. Chong, Y.-J. Chen, J.-C. A. Huang, P.-W. Chen and Y.-C. Tseng, RSC Adv. {\bf 8}, 7785 (2018).
	\bibitem{H.J.Wu} H.-J. Wu and W.-T. Yen, Acta Mater. {\bf 157}, 33 (2018).
	\bibitem{Dan} S. Dan, S. Kumar, S. Dan, D. Pal, S. Patil, A. Verma, S. Saha, K. Shimada, and S. Chatterjee, Appl. Phys. Lett. {\bf 120}, 022105 (2022).
	\bibitem{Steinberg} H. Steinberg, D. R. Gardner, Y. S. Lee, and P. Jarillo-Herrero, Nano Lett. {\bf 10}, 5032 (2010).
	\bibitem{Acosta} C. M. Acosta, E. Ogoshi, A. Fazzio, G. M. Dalpian, and A. Zunger, Matter {\bf 3}, 145 (2020).
	\bibitem{H.Yang} H. Yang, X. Peng, X. Wei, W. Liu, W. Zhu, D. Xiao, G. M. Stocks, and J. Zhong, Phys. Rev. B {\bf 86}, 155317 (2012).
	\bibitem{Z.Wang} Z. Wang and S. Law, Cryst. Growth Des. {\bf 21}, 6752  (2021).
	\bibitem{Kresse1} G. Kresse and J. Hafner, Phys. Rev. B {\bf 49}, 14251 (1994).
	\bibitem{Kresse2} G. Kresse and J. Furthm\"uller, Phys. Rev. B {\bf 54}, 11169 (1996).
	\bibitem{Blochl} Bl\"ochl, Phys. Rev. B {\bf 50}, 17953 (1994).
	\bibitem{Perdew} J. P. Perdew, K. Burke, and M. Ernzerhof, Phys. Rev. Lett. {\bf 77}, 3865 (1996).
	\bibitem{Grimme1} S. Grimme, J. Antony, S. Ehrlich, and H. Krieg, J. Chem. Phys. {\bf 132}, 154104 (2010).
	\bibitem{Grimme2} S. Grimme, S. Ehrlich, and L. Goerigk, J. Comput. Chem. {\bf 32}, 1456 (2011).
	\bibitem{Wolos} A. Wolos, S. Szyszko, A. Drabinska, M. Kaminska, S. G. Strzelecka, A. Hruban, A. Materna, M. Piersa, J. Borysiuk, K. Sobczak, and M. Konczykowski, Phys. Rev. B {\bf 93}, 155114 (2016).
	\bibitem{T.H.Wang} T. H. Wang and H. T. Jeng, arXiv:1608.00348 (2016).
		

	
\end{thebibliography}

\end{document}